# Neutralization of ion beam by electron injection, Part 1: Accumulation of cold electrons

C. Lan[1,2] and I. D. Kaganovich[2]

[1] *Institute of Fluid Physics, China Academy of Engineering Physics, Mianyang, 621900, P. R. China.*
[2] *Princeton Plasma Physics Laboratory, Princeton, New Jersey 08543, USA*

**Abstract:** Ion beam charge neutralization by electron injection is a complex kinetic process. Recent experiments show that resulting self-potential of the beam after neutralization by plasma could be much lower than the temperature of plasma electrons [Physics of Plasmas 23, 043113 (2016)], indicating that kinetic effects are important and may affect the neutralization of ion beam. We performed a numerical study of the charge neutralization process of an ion beam making use of a two-dimensional electrostatic particle-in-cell code. The results show that the process of charge neutralization by electron injection is comprised of two stages. During the first stage, the self-potential of the beam is higher than the temperature of injected electrons ($T_e/e$). Therefore, the neutralization of the ion beam is almost unaffected by $T_e$, and all injected electrons are captured by the ion beam. At the second stage, with the decline of the beam potential ($\varphi$), hot electrons escape from the ion beam, while cold electrons are slowly accumulated. As a result, $\varphi$ can be much lower than $T_e/e$. It is found in the simulations that during the accumulation of cold electrons, $\varphi$ scales as $\varphi \sim \sqrt{T_e}$. In addition, the results show that the transverse position of electron source has a great impact on ion beam neutralization. Slight shift of electron source leads to large increase of the beam potential because of increase in potential energy of injected electrons.

## 1 Introduction

Ion beams are widely used in many areas including accelerators [1], ion thrusters [2], inertial fusion, in particular fast ignition [3] and heavy ion fusion [4], and surface engineering [5-11], etc. In surface engineering, ion beams are usually used for etching, deposition of thin films and ion implantation. In addition, ion beams find applications in ion beam lithography and more recently nanopantography [5]. Many of these applications require that ion beam has an intense current at certain energy. Because of intense self-electric field of the ion beam, effective neutralization of space charge is necessary to prevent defocusing and decrease of ion flux, especially for low energy ion beams. For instance, as a typical application, heavy ion fusion driver like NDCX can generates tens of amperes of ion current at MeV but requires near-complete (99%) space charge neutralization in order to make ion beam focus [12-14]. In ion beam etching application, for example etching insulators, space charge compensation is particularly important for ion beams with high current density [6].

To compensate for a space charge potential of a positive ion beam, a sufficiently large number of electrons must be introduced from outside. This can be carried out by injecting electrons or producing plasma near or in the path of ion beam propagation, and it has been found that the neutralization degree of ion beam is associated with the scheme of introducing electrons [12]. One simple way to inject electrons is to use hot filaments, which can emit electrons from filaments [15, 16]. Temperature of emitted elections is equal to the temperature of filaments and is about 0.1 eV. For a plasma source operated in vacuum, the electron temperature is usually much higher than this value. In Ref. [13], the authors reported experiments on ion beam neutralization by a ferroelectric plasma source. This type of plasma source can provide 1~2 eV electrons by surface discharge. However, the experimental results show that the transverse electrostatic potential (~0.3 V) of neutralized ion beam is much lower than electron temperature, implying that the energy of the neutralizing electrons was below 0.3 eV. Such low temperature can be explained as follows. During the decay of space charge potential due to the ion beam during neutralization, the fast electrons can escape from formed potential well; while the cold electrons continuously accumulate inside the ion beam. This is similar to processes in textbook glow dc discharge leading to formation of negative glow with nearly room temperature electrons [17, 18].

In Ref. [19, 20], the authors pointed out that for a given stationary steady state beam, ultimate neutralization degree or beam potential is determined by electron temperature inside the ion beam. Here, beam potential is defined as the electric potential on the axis of the beam, it varies with neutralization degree. In the experiments, they studied the process of beam neutralization through residual gas ionization, and confirmed that Coulomb collisions of beam ions with neutralizing electrons provides an energy source for heating of captured electrons. However, the authors did not consider the influence of the way of introducing electrons on the behavior of electrons inside the ion beam. This is particularly important for neutralization of ion beams by hot filaments. Because in this case the injection of electrons is restricted to certain areas, the motion of neutralizing electrons inside the beam is nonlocal, e.g. they must move upstream or downstream to neutralize other places of the beam. As a result, the neutralization process and the behavior of neutralizing electrons are much more complicated compared with the neutralization through gas ionization. Accumulation and movement of electrons in potential well formed by the beam is complicated and resulting velocity distribution function of neutralizing electrons is anisotropic and far from Maxwellian distribution. Moreover, in the neutralization process nonlinear electrostatic solitary waves (ESWs [21-27]) can be excited [28].

In this work, we employ an implicit 2D particle-in-cell (PIC) code to investigate numerically the ion beam neutralization by filaments. In this part 1 paper, we focus mainly on the accumulation process of cold electrons and its influence on the neutralization of ion beams. The detailed description of the resulting excitation of ESWs during the accumulation of cold electrons is presented as a companion part 2 paper to this article.

The paper is organized as follows. In the second section, a 2D simulation model of ion beam neutralization by injecting electron is described. The simulation results and corresponding discussion are in section 3. We will show how cold electrons are accumulated in the potential well of ion beam and study how beam potential scales with the temperature of injected electrons. In addition, the effect of initial spatial distribution of injected electrons, for instance, the position of electron source, are also discussed. Finally, conclusions are summarized in Section 4.

**2 Simulation model**

The 2D set up shown in Fig. 1 was used to simulate the ion beam transport in a metal pipe. Electrons are injected on the axis to neutralize the ion beam. Ion beam, electron injection and transporting metal pipe comprise a simple but complete physical model of neutralization. Such an electron injection scheme represents the electron emission by hot filaments placed into the beam path [15]. For simplicity, the model is 2D in *x-y* uniform Cartesian coordinate system, where *x* is in the direction of the ion beam propagation and *y* is the transverse direction. The size of computational domain is 40 cm × 6 cm. The cell size of uniform Cartesian grid is 0.25 mm, which leads to a grid of 1600 × 120 cells. Monoenergetic Ar$^+$ beam with energy $E_b$=38 keV and initial density $n_b = 1.75 \times 10^{14}$ m$^{-3}$ are injected from the left boundary. The parameters of ion beam are chosen close to those of Princeton Advanced Test Stand at PPPL [13]. Because ion beam current is very low and ion beam flow velocity $V_b$ satisfies $V_b \ll c$, where *c* is the light speed, the inductive magnetic field of ion beam in vacuum can be neglected compared to its self-electric field, and the system can be treated electrostatically. Therefore, in 2D PIC simulations, Poisson's equation was used to obtain the electric potential from charge density.

Upper and lower metal walls are totally absorbing boundaries for particles. For left and right boundaries, considering that in experiments ion beams are usually extracted through a metal grid, and collected by a Faraday cup or directly hit a metal target after traveling a distance, so both left and right walls of the model can be treated as metal boundaries for electric field and absorbing boundaries for particles. When ions hit a metal wall, a great number of secondary electrons will be created. However, these electrons will oscillate many times in the potential well formed by the ion beam. As a result, thermalizing and cooling these electrons require a much longer time scale. So, for simplicity ion induced secondary electron emission was not considered in our simulations.

The collisions between charged particles and neutral particles were not modeled and Coulomb collisions between charged particles were also neglected as they only weakly affect the neutralization process. The time step of the simulations was 80 ps, and single simulation lasted for more than 30 μs. 6000 particles per cell were used to reduce numerical noise. Simulations were run with 40 cores on the Princeton University Adroit supercomputer.

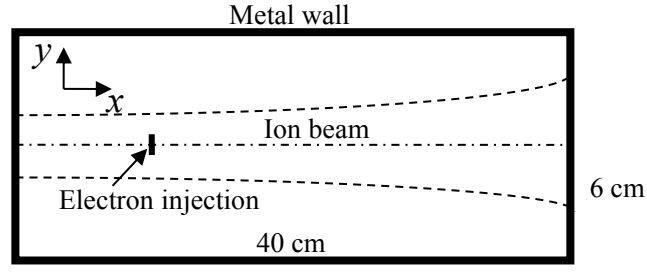

Fig. 1. Schematic of the simulation model. The model is 2D in *x-y* Cartesian coordinates system. Ion beam moves along the axis of the model with speed $V_b$. Electrons are injected on the axis. Dashed curves represent the envelope of ion beam.

## 3 Results and discussion
### 3.1 Accumulation of cold electrons during ion beam neutralization

The ion beam is uniformly injected into the initially empty domain from the center of left wall with fixed ion current and 5 mm beam width. After propagating for about 1 μs, the ion beam reaches the right wall ($V_b$ =37.5 cm/μs). Then electrons start to be injected at 1 μs. In order to inhibit excitation of various plasma waves or instabilities, injected electron current should be sufficiently small, with 1/3 of ion beam current in our simulation set up. The position of the injection is on the path of the ion beam (*x*=10 cm). To study a wider range of physical laws, electron temperature is not limited to the typical temperature of hot filaments, and varied within several electron volts. The injected ion beam is perfectly collimated with zero divergence, so that the expansion of the ion beam is only determined by the remaining space charge and the beam envelope *R(x)* is described by the perveance *Q* of the ion beam:

$$\frac{\partial^2 R}{\partial x^2} = \frac{(1-f)Q}{R}, \qquad (1)$$

where *f* is the neutralization fraction of the ion beam.

Figure 2 shows the evolution of particle densities profiles during the neutralization of the ion beam. As seen from this figure, the initially expanding ion beam gradually shrinks in envelope after the neutralization begins. According to the downstream divergence angle and the width of the ion beam, one can calculate from Eq. 1 that the neutralization fraction has reached more than 99% at *t*=4.3 μs. Because of very deep potential well, electrons initially are distributed near the axis of the ion beam. As the neutralization fraction increases, electrons gradually fill up the whole ion beam and some of hot electrons escape the potential well and get lost on the walls.

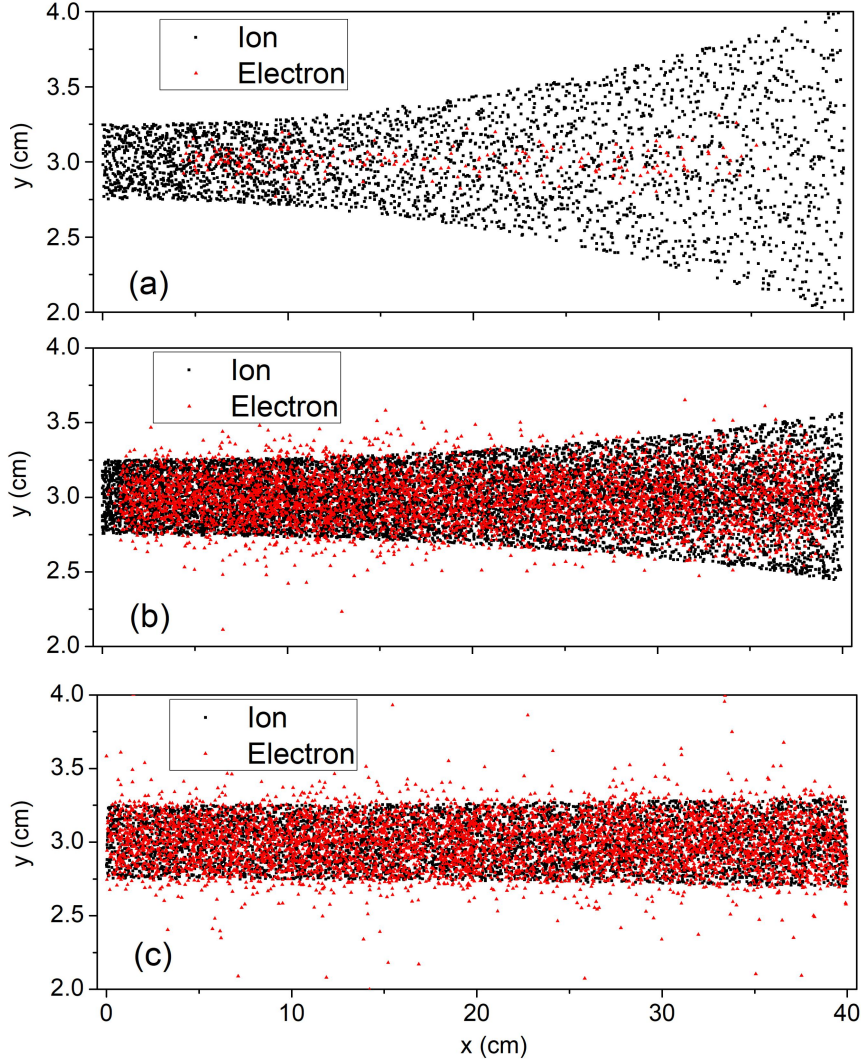

Fig. 2. Spatial distributions of electrons and ions at (a) $t$=1.2 μs, (b) $t$=3.2 μs and (c) $t$=4.3 μs. The temperature of injected electrons is $T_e$=2 eV.

If we do not consider the loss of hot electrons, the time required to completely neutralize ions is determined by

$$\Delta t = \frac{I_i L}{I_e} \sqrt{\frac{2m_i}{E_b}}, \qquad (2)$$

where $I_i$ and $I_e$ are the currents of ion beam and injected electrons, respectively. $L$ is the length of ion beam and $m_i$ is ion mass. In this simulation, we have $I_i/I_e$=3, therefore $\Delta t$=3.2 μs. However, because of the loss of hot electrons the time required to neutralize ions is longer than this value.

Figure 3 plots temporal evolution of the beam potential on the axis and the neutralization fraction of the ion beam. The ion beam potential reaches a steady state at around 1 μs. As electrons are continuously injected into the ion beam, the beam potential is decreased almost linearly from a maximal value of about 220 V. After around $\Delta t$=2.8 μs, i.e. $t$≈3.8 μs, the beam potential curve as function of time reaches an inflexion point. At this point, attainable neutralization fraction is 98% and the beam potential is about 4 V, which is twice of the temperature of injected electrons, $T_e$=2eV for this case. After that, the neutralization of the ion beam enters second stage. Hot electrons with energies larger than the beam potential could escape from the potential well, meanwhile more and more cold electrons are being trapped in this potential well, causing the beam potential to continue to decline. As a result, minimum residual potential of the ion beam is far below the temperature of injected electrons, as shown in Fig. 3. This is a very slow dynamic process. After this stage, the neutralization fraction is increased from 98% to more than 99%. As the beam potential decreases, electrons inside the ion beam become colder and colder, until reaching a particle balance that the rate of capture of cold electrons is equal to the rate at which cold electrons are heated and escaped.

In Fig. 3 we can see that the second stage of neutralization is very important. Without this stage, 99% neutralization fraction is hard to obtain. For the case of $T_e/e\varphi_0$>1%, where $T_e$ is the temperature of injected electrons and $\varphi_0$ is the beam potential before neutralization, in order to get over 99% neutralization fraction, the accumulation of cold electrons is necessary.

It should be noted that the occurrence of the second stage of neutralization does not rely on the single point electron injection, and even does not rely on the injection location of electrons (see Fig. 10 below). As long as electrons are continuously generated near the potential well of the ion beam, the accumulation of cold electrons will occur.

The ultimate residual potential of the ion beam is determined by many factors such as Coulomb collisions between charge particles (not included in this model) [19], wave-particle interaction, the influence of electron source and numerical heating caused by macro-particle and spatial gridding of the PIC method [29], etc. The study of these heating mechanisms of cold electrons is beyond the scope of this paper. We focus our attention to the behavior of cold electrons before reaching the particle balance. It can be seen in Fig. 3 that the second stage, i.e. accumulation of cold electrons, can last for more than 15 μs. The residual potential of the ion beam we can obtain from this simulation is close to 0.7 V, which is about twice of measured residual potential in Ref. [13]. The difference may be due to the different ways in which electrons are injected and numerical heating that does not exist in experiments.

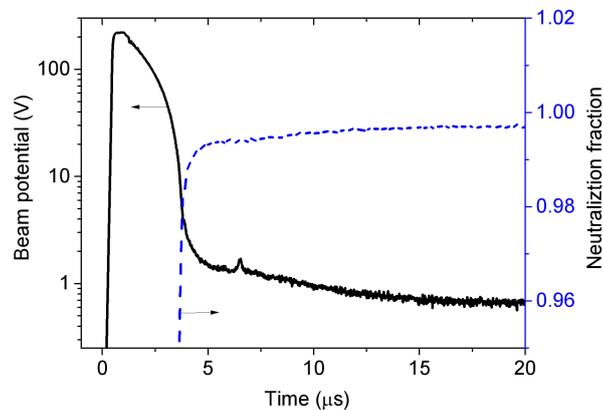

Fig. 3. Temporal evolutions of the beam potential at *x*=20 cm and the neutralization fraction of the ion beam. Beam potential is defined as the potential on the axis of the beam.

Because the temperature of cold electrons is finite and the neutralization degree of the ion beam is close to 1, some of cold electrons are located outside the ion beam. Consequently, the distribution of electrons outside the ion beam leads to the generation of radial electric field. Figure 4 shows the distributions of particle densities in the *y* direction at different moments. At the beginning of neutralization, all the injected electrons are located inside the ion core. When the beam potential is reduced to several volts, a large number of hot electrons start spilling out. As electrons captured by the ion beam get colder, the number of electrons distributed outside the ion beam becomes smaller.

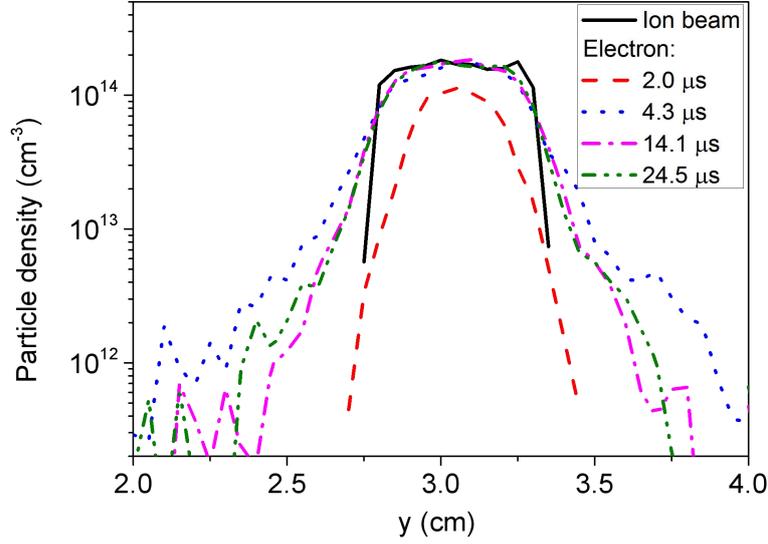

Fig. 4. Distributions of the electron density and the ion density along the *y* direction at different moments (*x*=20 cm).

The profile of potential along the *y* direction at different moments of time are shown in Fig. 5. Figure 4(a) and (b) respectively correspond to the first and the second stages of neutralization. We can see clearly how potential well evolves in the whole process of neutralization. Because of 2D model, at the first stage the potential outside the ion beam is linearly decreased with the transverse distance from the ion beam and linearly decreasing with time. However, at the second stage because of the distribution of electrons outside the ion beam, actual depth of potential well is larger than the beam potential, implying that the kinetic energy of cold electrons can be larger than the beam potential at that moment. Furthermore, the more cold electrons accumulate, the greater the deviation. This result does not contradict the conclusion of Ref. [13], because of different ways of introducing electrons. In the experimental studies in Ref. [13], electrons were generated on the surface of outer wall. But here electrons are directly injected into the center of the ion beam.

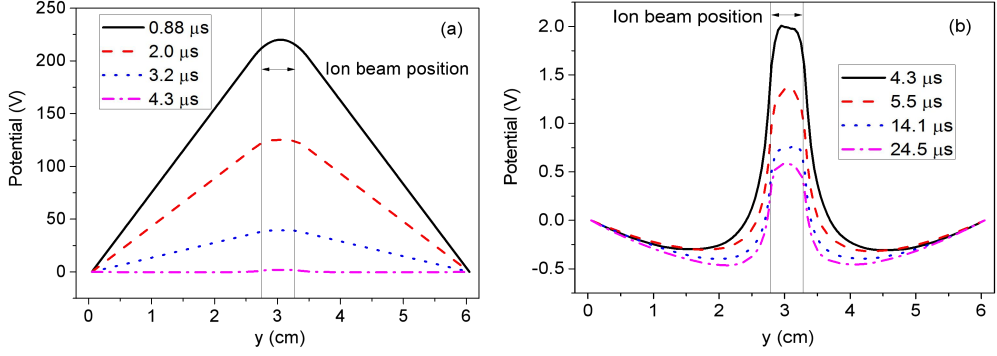

Fig. 5. Distribution of the potential along the *y* direction (*x*=20 cm) at different moments, (a) *t* is from 0.88 μs to 4.3 μs and (b) *t* is from 4.3 μs to 24.5μs.

The accumulation of cold electrons and the depletion of hot electrons can be clearly seen through the evolution of electron velocity distribution functions (EVDFs), which are presented in Fig. 6. About 4 μs later, hot electrons begin to escape. But the accumulation of cold electrons has been going on since the beginning of electron injection. Due to the fact that the scale of potential well formed by the ion beam varies greatly in different directions, the behavior of neutralizing electrons exhibits anisotropy during neutralization. When electrons are injected into the center of ion beam, cold electrons tend to drift along the *x* direction under the action of longitudinal electric field. This longitudinal electric field is caused by the space charge near the location of electron injection. If the energies of cold electrons are below than the beam potential, they will bounce back into the ion beam from the left and right walls, forming two streams of cold electrons moving in opposite direction with comparable densities, as shown in Fig. 6(a). However, this phenomenon is not observed in the *y* direction. The EV$_y$DF is basically close to Maxwellian distribution in the range of low energy (<0.5 eV). The reason may be due to much smaller scale of the potential well in the *y* direction (~1 cm in the *y* direction vs. 40 cm in the *x* direction). As a result, transverse motion of cold electrons are effectively restrained.

The double-peak EV$_x$DF shown in Fig. 6(a) indicates that the plasma composed of the ion beam and neutralizing electrons is unstable. Subsequent evolution will be discussed in the last part of this section. As more and more cold electrons accumulate in the potential well, we see that the double-peak EV$_x$DF gradually disappear and the EV$_x$DF tends to be Maxwellian.

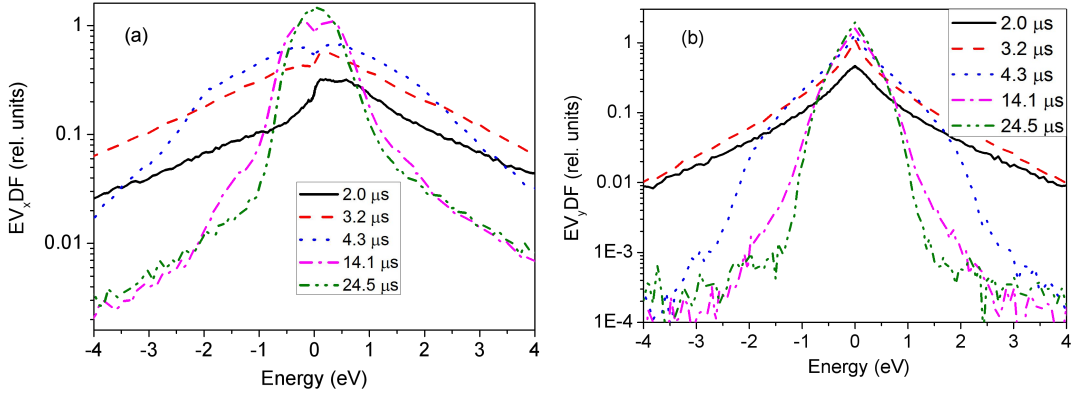

Fig. 6. Temporal evolutions of $EV_xDF$ and $EV_yDF$. Electron velocity is represented by its energy. All electrons in the domain are taken into account, but electrons outside the ion beam take up a very small portion. The asymmetry of $EV_xDF$ is caused by the spatial asymmetry of electron injection.

### 3.2 Scaling of beam potential with the temperature of injected electrons

In Ref. [19], it has been found experimentally that the measured potential drop $\Delta\varphi$ from the center of the beam to the beam periphery has the following relation,

$$\Delta\varphi = C\sqrt{\Delta\varphi_0 T_e/e}, \qquad (3)$$

where $\Delta\varphi_0$ is the potential drop of unneutralized ion beam, $C$ is a coefficient whose value depends on the beam profile and applied electron source. $T_e$ is the temperature of hot emitter, which is usually assumed to be equal to the temperature of emitted electrons. For the correlation between $\Delta\varphi$ (or beam potential $\varphi$, with a difference in coefficient for a given profile of ion beam) and $T_e$, we still lack sufficient numerical study [12]. In this section, simulations were carried out to test the validity of this scaling relation.

In simulations $T_e$ was changed from 2 eV to 6 eV, while other parameters kept the same. Fig. 7 shows how the beam potential changes with the temperature of injected electrons. For the big hump appeared in beam potential when $T_e$=6 eV, it is caused by the ESWs, which will be discussed in the Part 2 paper. Because $T_e$ is still far smaller than initial beam potential, we see in Fig. 7 that $T_e$ does not affect the neutralization of first stage. At the second stage of neutralization, as expected, higher electron temperature leads to slower accumulation of cold electrons.

The beam potentials at arbitrary three moments of the second stage are plotted with respect to $\sqrt{T_e}$ in Fig. 8. A near linear correlation between the beam potential and $\sqrt{T_e}$ can be clearly seen. As per Equation (3), the beam potential should scale as $\varphi \sim \sqrt{T_e}$. This is demonstrated in Fig. 8 and consistent with experiment observations

[19].

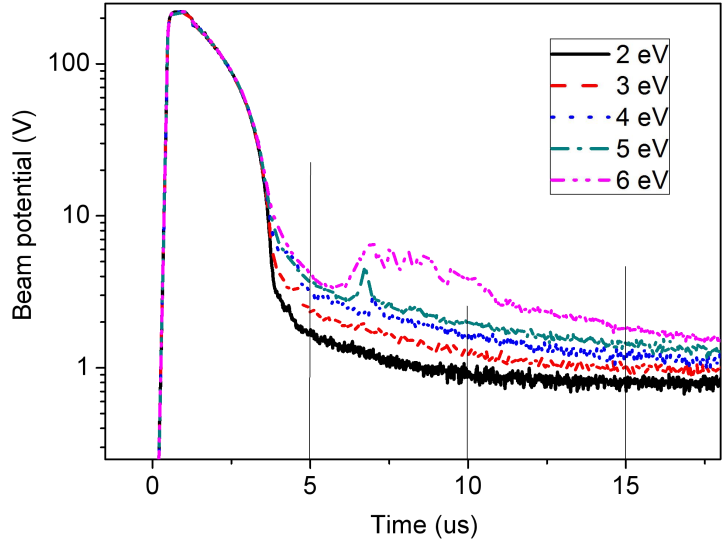

Fig.7. Temporal evolutions of the beam potential for different temperatures of injected electrons

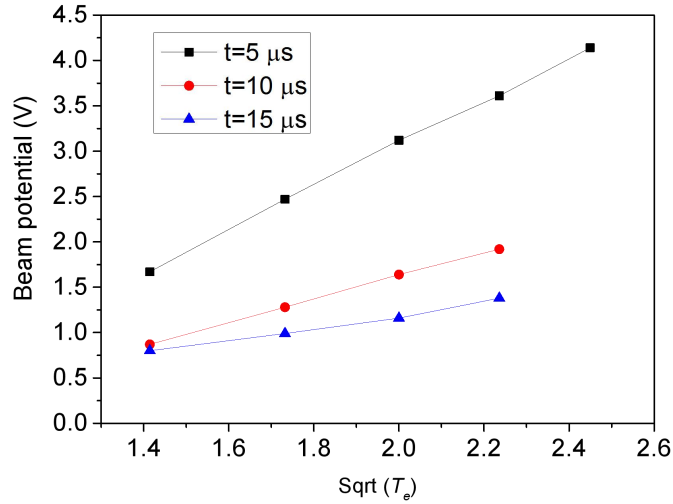

Fig. 8. Scaling relation between the beam potential and the temperature of injected electrons at different moments.

**3.3 Influence of injection position on ion beam neutralization**

In previous numerical simulations, electrons were injected on the path of the ion beam, i.e. the lowest point of potential well of the ion beam. In this case, neutralizing electrons can gradually fill this potential well until they start to escape and collective processes for transverse electron motion is not observed. Nevertheless, if electron injection is shifted from the axis of the ion beam, the neutralization process of the ion beam will be quite different. In order to clearly show the effect of injection position, we positioned the electron source far away ($y=0$) from the ion beam, as shown in Fig. 9. In this case, we see that besides the motion in the beam propagation direction, electrons emitted from the box edge experience great transverse oscillation around the ion beam. Due to higher transverse acceleration, electrons constantly overshot the ion beam center and bounce between two edges, thus decreasing their residence time within the ion beam. Some electrons are thermalized through electron-electron two-stream instability and reside in the ion beam, but this process is very slow. Consequently, the neutralization degree of the ion beam is very low and the self-consistent electric field causes the ion beam to undergo defocusing.

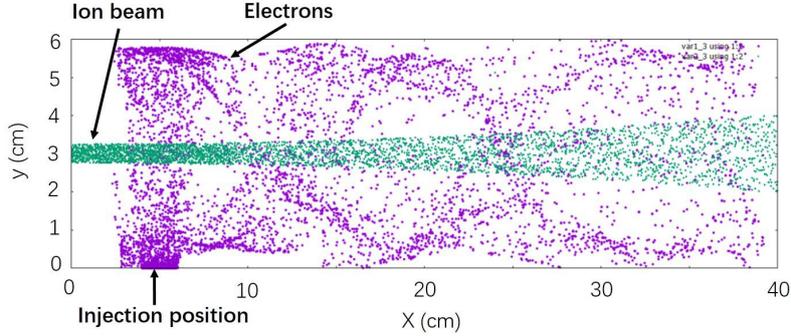

Fig. 9. Oscillation of neutralizing electrons around the ion beam when electron injection is shifted from the beam axis ($t$=1.28 μs or 177 ns after electron injection).

We varied the position of electron source in simulation to investigate its influence on ion beam neutralization. Figure 10 shows the temporal evolution of the beam potential during ion beam neutralization, three curves corresponding to three positions of electron injection. It is evidently seen that slight alteration of electron source in the $y$ direction (from 0 to 5 mm) leads to an order of magnitude increase in beam self-potential, thereby resulting in much higher beam potential than the temperature of injected electrons. So the neutralization of the ion beam is very sensitive to the transverse position of electron source. It is worth noting that the two-stage neutralization process mentioned above still exists for an electron source shifted to the periphery, and the turning point between two stages almost keeps unchanged. Meanwhile, in logarithmic coordinate slop of decline after 7.5 μs is nearly the same for relatively larger shift of electron source, indicating that even though very slow, the accumulation of electrons inside the ion beam causes exponential decline of the beam self-potential.

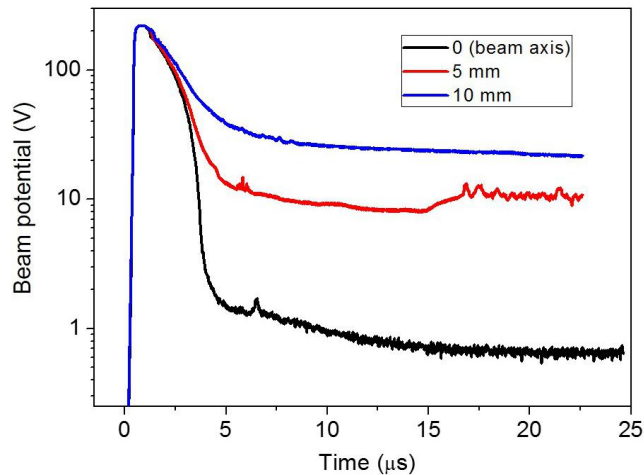

Fig. 10. Influence of the position of electron injection on the beam potential during ion beam neutralization. 5 mm and 10 mm are distances from the beam axis.

## 4 Conclusions

In this paper, we have presented 2D particle-in-cell numerical studies of the neutralization of an ion beam by electron injection. The simulation results show that the neutralization process is comprised of two stages. In the first stage, due to high potential well, almost all electrons can be captured by ion beam, leading to rapid decline of the beam self-potential. In the second stage, cold electrons are slowly accumulated while hot electrons with energies higher than the beam potential are escaped from potential well. The second stage takes much longer time than the first stage and resulting beam self-potential becomes far less than the electron temperature of the source. Without the second stage of neutralization, the neutralization degree will not exceed 99%. Our numerical simulations confirm that the residual potential linearly scales as $\sqrt{T_e}$ at different moments during the accumulation of cold electrons, where $T_e$ is the temperature of injected electrons. In the process of cold electron accumulation, because there is a big difference in the size of potential well in both directions, the electron distribution function exhibits some anisotropy. Double-peak EV$_x$DF appeared at the second stage of neutralization indicates that cold electron EVDFs is unstable due to the two-stream instability. In addition, the neutralization process of ion beam is very sensitive to the transverse position of electron source. Transverse shift of electron source from the center to the periphery causes cold electrons to accumulate more slowly and reduce the neutralization degree of the ion beam.

Acknowledgments: The work of Igor D. Kaganovich was supported the by U.S. Department of Energy. The work of Chaohui Lan was supported by the International Cooperation and Exchange Fund of China Academy of Engineering Physics.

References
[1] I. Blumenfeld *et al.*, Nature (London) **445**, 741 (2007); P. Chen *et al.*, Phys. Rev. Lett. **54**, 693 (1985); R. Govil *et al.*, Phys. Rev. Lett. **83**, 3202 (1999).
[2] S. Humphries, Appl. Phys. Lett. 32, 792 (1978); S. Humphries *et al.*, Phys. Rev. Lett. **16**, 995 (1981).
[3] A. J. Kemp *et al.*, Phys. Rev. Lett. **97**, 235001 (2006); R. J. Mason, Phys. Rev. Lett. **96**, 035001 (2006).
[4] P. K. Roy *et al.*, Phys. Rev. Lett. **95**, 234801 (2005); B. G. Logan *et al.*, Nucl. Instrum. Methods Phys. Res., Sect. A **577**, 1 (2007); I. D. Kaganovich *et al.*, *ibid.* **577**, 93 (2007); A. B. Sefkow *et al.*, *ibid.* **577**, 289 (2007); D. R. Welch, *ibid.* **577**, 231 (2007).
[5] J. P. Chang, J. C. Arnold, G. C. H. Zau, H.-S. Shin and H. H. Sawin, Journal of Vacuum Science & Technology A: Vacuum, Surfaces, and Films **15**, 1853-1863 (1997).
[6] A. Stojkovic, M. Radmilovic-Radenovic, Z. Lj. Petrovic, Material Science Forum **494**, 297-302 (2005).

[7] M. Watanabe, D. M. Shaw, and G. J. Collins, Applied Physics Letters **79**, 2698 (2001).

[8] J. Ishikawa, H. Tsuji, K. Shibutani, H. Ikai, Y. Gotoh, Proceeding of 1998 International Conference on Ion Implantation Technology, 716-719 (1998).

[9] C. Fusellier, L. Wartski, J. Aubert, C. Schwebel, Ph. Coste, and A. Chabrier, Review of Scientific Instruments **69**, 1153 (1998).

[10] I. G. Brown and J. Washburn, Nucl. Instrum. Methods Phys. Res., Sect. B **21**, 201 (1987).

[11] L. Dumas, E. Quesnel, J. -Y. Robic, Y. Pauleau, Thin Solid Films **382**, 61 (2001).

[12] I. D. Kaganovich, R. C. Davidson, M. A. Dorf, E. A. Startsev, A. B. Sefkow, E. P. Lee, and A. Friedman, Phys. Plasmas **17**, 056703 (2010).

[13] A. D. Stepanov, E. P. Gilson, L. R. Grisham, I. D. Kaganovich, and R. C. Davidson, Phys. Plasmas **23**, 043113 (2016).

[14] A. D. Stepanov, J. J. Barnard, A. Friedman, E. P. Gilson, D. P. Grote, Q. Ji, I. D. Kaganovich, A. Persaud, P. A. Seidl, T. Schenkel, Matter and Radiation at Extremes **3**, 78-84 (2018).

[15] D. V. Rose, D. R. Welch, S. A. MacLaren, Proceedings of the 2001 Particle Accelerator Conference, 3003 (2001).

[16] S. A. MacLaren, A. Faltens and P. A. Seidl, Phys. Plasmas **9**, 1712 (2002).

[17] L. D. Tsendin, Plasma Sources Sci. Technol. **12**, s51-s56 (2003).

[18] V. A. Rozhansky and L. D. Tsendin, *Transport Phenomena in Partially Ionized Plasma* (CRC Press, 2001).

[19] M. D. Gabovich, I. A. Soloshenko, and A. A. Ovcharenko, Ukr. Fiz. Zh. Russ. Ed. **15**, 934 (1971).

[20] I. A. Soloshenko, Review of Scientific Instruments 67, 1646 (1996).

[21] R. L. Morse and C. W. Nielson, Phys. Rev. Lett. **23**, 1087 (1969).

[22] H. L. Berk, C. E. Nielsen, and K.V. Roberts, Phys. Fluids **13**, 980 (1967).

[23] H. Schamel, Phys. Scr. **20**, 336 (1979).

[24] H. Schamel, Plasma Phys. **13**, 491 (1971); **14**, 905 (1972).

[25] H. Schamel, Phys. Plasmas **7**, 4831 (2000).

[26] I. H. Hutchinson, Phys. Plasmas **24**, 055601 (2017).

[27] H. Matsumoto, H. Kojima, T. Miyatake, Y. Omura, M. Okada, I. Nagano, and M. Tsutsui, Geophys. Res. Lett. **21**, 2915 (1994).

[28] C. Lan and I. D. Kaganovich, Phys. Plasmas. **26**, 050704 (2019).

[29] H. Ueda, Y. Omura, H. Matsumoto and T. Okuzawa, Computer Physics Communications **79**, 249-259 (1994).